\documentstyle[iopconf1]{article}

\input epsf

\begin{document}

\title{Direct Detection of Nonbaryonic Dark Matter}

\author{Laura\ Baudis \footnote{E-mail: Laura.Baudis@mpi-hd.mpg.de}, \\
H.\ V.\ Klapdor-Kleingrothaus \footnote{E-mail: 
          klapdor@daniel.mpi-hd.mpg.}}

\affil{
Max--Planck--Institut f\"ur Kernphysik, \\
P.O.Box 10 39 80\\ D--69029 Heidelberg, Germany}

\beginabstract 
Weakly Interacting Massive Particles (WIMPs) are leading candidates 
for the dominant part of the mass density of the Universe. 
Here we will review direct WIMP detection techniques by giving 
examples of currently running experiments, and 
present the status of the most promising future projects.
\endabstract

\section{Introduction}

There is strong observational and theoretical
evidence that dark, nonbaryonic matter accounts 
for about a third of the critical density of the Universe (for a recent 
review see \cite{mturner} and references therein).
Many candidates have been proposed  
and some of them (cosmions, heavy Dirac neutrinos) have already
been rejected. Slow thermal relics born in an early phase 
of the Universe, stable or  very long lived, are excellent  
candidates for nonbaryonic dark matter \cite{kolbturner}. These weakly interacting 
($\sigma \leq \sigma_{weak})$, massive (1 GeV - 1 TeV) particles
(WIMPs) arise independently from cosmological considerations 
in supersymmetric models as neutralinos - the lightest
supersymmetric particles. Direct detection of neutralinos can occur 
in very low background experiments, where the elastic neutralino scattering 
off target nuclei \cite{goodwitt} is exploited.

In the following, after giving a short overview 
on the principles of direct detection,
we will describe existing or planned 
experimental techniques aiming at direct WIMP detection, 
illustrated by examples of currently running experiments and of 
future projects. We do not intend a complete coverage of all  
direct detection experiments, but rather present some of the most
promising ones.

\section{Principles of direct detection}

The differential rate for WIMP elastic scattering off nuclei is given by 
\cite{smith_lewin96}

\begin{equation}
\frac {dR}{dE_R}=N_{T}
\frac{\rho_{0}}{m_{W}}
                    \int_{v_{\rm min}}^{v_{\rm max}} \,d \vec{v}\,f(\vec v)\,v
                     \,\frac{d\sigma}{d E_R}\, , 
\label{eq1}
\end{equation}

where $N_T$ represents the number of the target nuclei,
$m_W$ is the WIMP mass and $\rho_0$
the local WIMP density in the galactic halo,
$\vec v$ and $f(\vec v)$ are the WIMP
velocity and velocity distribution function  in the Earth frame
 and ${d\sigma}/{d E_R}$ is the WIMP-nucleus differential cross section.

The nuclear recoil energy is given by
$E_R={{m_{\rm r}^2}}v^2(1-\cos \theta)/{m_N}$,
where $\theta$ is the  scattering
angle in the WIMP-nucleus center-of-mass frame,
$m_N$ is the nuclear mass and $m_{\rm r}$ is the WIMP-nucleus
reduced mass. The velocity $v_{\rm min}$ is defined as 
$v_{\rm min} = (m_N E_{th}/2m_{\rm r}^2)^{\frac{1}{2}}$, where $E_{th}$
is the energy threshold of the detector, and 
$v_{\rm max}$ is the 
escape WIMP velocity in the Earth frame.  

The standard lore assumes a Maxwell-Boltzmann distribution for the
WIMP velocity in the galactic rest frame, with a velocity dispersion 
of v$_{rms} \approx$ 270 km s$^{-1}$ and an escape 
velocity of v$_{esc} \approx$ 650 km s$^{-1}$.

The differential WIMP-nucleus cross section can be divided into two 
separate parts: an effective scalar coupling between the WIMP and the 
nucleus (proportional to A$^2$, where A is the target atomic mass) and 
an effective coupling between the spin of the WIMP and the total spin of the nucleus.
In general the coherent part dominates the interaction (for
neutralinos) and the cross section 
can be factorized as 
$\frac{d\sigma}{d E_R} \propto {\sigma_0} F^2(E_R)$, where 
 $\sigma_0$ is the point-like scalar WIMP-nucleus
cross section and $F(E_R)$ denotes the nuclear form factor, 
expressed as a function of the recoil energy.

\begin{figure}[t]  
\centering 
\leavevmode\epsfxsize=300pt  
\epsfbox{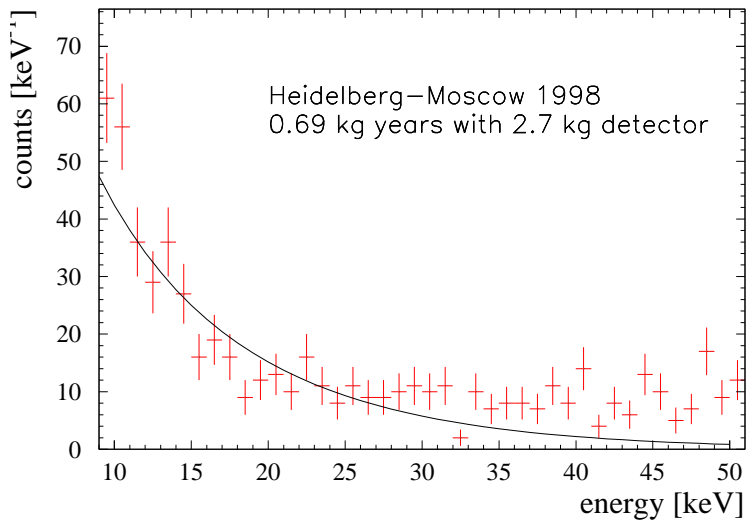}
\caption{\label{ang2} Low energy spectrum of one of the enriched
  $^{76}$Ge detectors of the Heidelberg-Moscow experiment and a
  theoretical spectrum for a 100 GeV WIMP mass.} 
\end{figure}

The left side of equation \ref{eq1} is the measured quantity  in a detector,
the right side represents a theoretical
WIMP spectrum (see Figure \ref{ang2} for an example of a measured spectrum and a calculated 
WIMP spectrum for m$_W$ = 100 GeV). 
It includes the WIMP properties which are completely unknown,
like the WIMP mass $m_W$ and elastic cross section $\sigma_0$, 
quantities accessible from astrophysics, like the density of WIMPs in the 
halo, $\rho_0$, the WIMP velocity distribution and the escape
velocity (which are however prone to large uncertainties) and
detector specific parameters, like mass of target nucleus, energy
threshold and nuclear form factor.

It is straightforward to see that 
a WIMP with a typical mass between a few GeV and 1 TeV 
will deposit a recoil energy below 100 keV in a terrestrial detector.
As for the predicted event rates, scans of the MSSM parameter 
space under additional assumptions  (GUT, mSUGRA, etc) and accounting for
 accelerator and cosmological constraints, yield  about 10$^{-5}$ to 10
events per kilogram detector material and day \cite{theo_rates}.

Evidently, in order to be able to measure a WIMP spectrum, 
low energy threshold, low background and high mass detectors are 
required.

In such a WIMP detector, the recoil energy of the scattered nucleus 
is transformed into
a measurable signal, like ionization, scintillation or lattice
excitations and at least one of the above quantities is detected.

An additional information is given by the pulse shape of the interaction, 
which  can lead to background reduction in some cases (see Section 
\ref{existing}).

A low background can be achieved in both passive and active ways.
Passive methods range from high material selection of detector
components to various specific shieldings against the natural
radioactivity of the environment and against cosmic rays and 
secondary particles produced in cosmic ray interactions.
Active background reduction implies either an active shielding
surrounding the detector (against gammas and/or muons) 
and/or the recording of  two of the above mentioned quantities 
(ionization and phonon, scintillation light and phonon or 
ionization and scintillation signals), since in general the amount of energy shared
between two  observables is different for nuclear recoils
and photon induced interactions. Consequently, a discrimination
between WIMPs (or neutron) induced recoils and gamma- or electron
induced events becomes possible.

Identifying the measured spectrum as a WIMP induced one is by far
not trivial, since both background (radioactive and noise) and WIMP
signal are exponentially decreasing shaped. 
For this purpose an additional quantity, a WIMP signature
is requested.  

The Earth`s motion through the galaxy induces both a seasonal variation of 
the total event rate \cite{fre86,fre88} and a forward-backward 
asymmetry in a directional signal \cite{spergel88,copi99}. 

The annual modulation of the WIMP signal arises due to the Earth motion 
in the galactic frame, which is a superposition of the Earth rotation
around the Sun with that of the Sun around the galactic center:

\begin{equation}
v_E = v_\odot + v_{\rm orb} \; \cos \gamma \; \cos \omega (t - t_0), 
\end{equation}

where $v_\odot = v_0$ + 12 km s$^{-1}$ ($v_0$ $\approx$ 220  km s$^{-1}$),  
$v_{\rm orb}$ $\approx$ 30 km s$^{-1}$ denotes 
the Earth orbital speed around the Sun, the angle 
$\gamma \approx 60^0$ is the inclination of the Earth orbital plane 
with respect to the galactic plane and $\omega = 2 \pi / 1 \mbox{yr}$, 
$t_0 =$ June  2$^{\rm nd}$.  

The expected time dependence of the count rate can  be approximated 
by a cosine function with a period of T = 1 year and a phase of $t_0
=$ June  2$^{\rm nd}$:

\begin{equation}
S(t) = S_0 + S_m cos \omega (t - t_0),
\end{equation}

where $S_0$, $S_m$ are the constant and the modulated amplitude of the signal, 
respectively.
In reality, an additional contribution to S(t) from the
background (B) must be considered,  which is supposed to  be constant in time.

The expected seasonal modulation effect is very small 
(of the order of $v_{\rm orb}$/$v_0 \simeq $ 0.07), requiring
large masses and/or large counting times and an excellent long-term stability 
of the experiment.

A much stronger signature would be given by the ability to detect the axis and 
direction of the recoil nucleus.
In \cite{spergel88} it has been shown, that the WIMP interaction rate as a function 
of recoil energy and angle $\theta$ between the WIMP velocity and
recoil direction 
(in the galactic frame) is:

\begin{equation}
\frac{d^2R}{dE_R d cos\theta} \propto \rm{exp}\left[-\frac{(v_\odot cos\theta - v_{min})^2}{v_0^2} \right],  
\end{equation}

where v$_{min}^2$ = (m$_N$ + m$_W$)$^2$E$_R$/2m$_N$m$_W^2$ 
and v$_0^2$=3v$_\odot^2$/2.

The forward-backward asymmetry yields thus a large effect of the order of 
$\mathcal{O}$(v$_\odot$/v$_0$)$\approx$ 1 and much less events are needed 
to discover a WIMP signal than in the case of the seasonal modulation \cite{copi99}.
 

\section{Existing experiments}
\label{existing}

After the seminal paper of Goodman$\&$Witten in 1985 \cite{goodwitt},
first limits on WIMP-nucleon cross sections were derived from at 
that time already existing Germanium double beta decay experiments \cite{ge_exp}.
Since then, more then a dozen of dedicated WIMP dark matter
experiments were built up and are delivering data, while even more 
are planned for the future (see for example \cite{dark98,la98}).

Dark matter experiments can be classified by the employed technique 
\cite{caldwell,morales} or by their ability to distinguish between
nuclear recoils and other types of interactions \cite{yorckdm}.

Another type of classification would be by the
ability of the experiments to detect a WIMP signature, which
however is basically equivalent (with one
exception,  i.e. DAMA) to our subdivision  in currently 
running  experiments and in future projects. 
The annual modulation signature can in principle be looked for 
in any kind of experiment. However, as it has been shown
\cite{yorckmod,hasblg,cebrian}, there exist statistical
limitations for extracting a periodic signal  from the data  for low
background experiments and depending on the signal/noise ratio 
a minimal target mass is required. Consequently, only high mass
experiments are well suited and will have a realistic chance 
to detect a seasonal variation of the WIMP rate with high statistical 
significance.

To detect a forward-backward asymmetry, a directional sensitive
detector is required. Because of the inherent difficulties in 
conceiving such a detector, only one realistic project exists up to now 
(the DRIFT project, see section~\ref{future}).

Low mass experiments with no directional information 
can set stringent limits on the WIMP-nucleon cross section and thus
test (and exclude) supersymmetric models and eventual ``evidence
regions'' in SUSY parameter spaces found by high mass experiments,
under the condition that their radioactive background is low enough.

\begin{figure}
\centering 
\leavevmode\epsfxsize=330pt  
\epsfbox{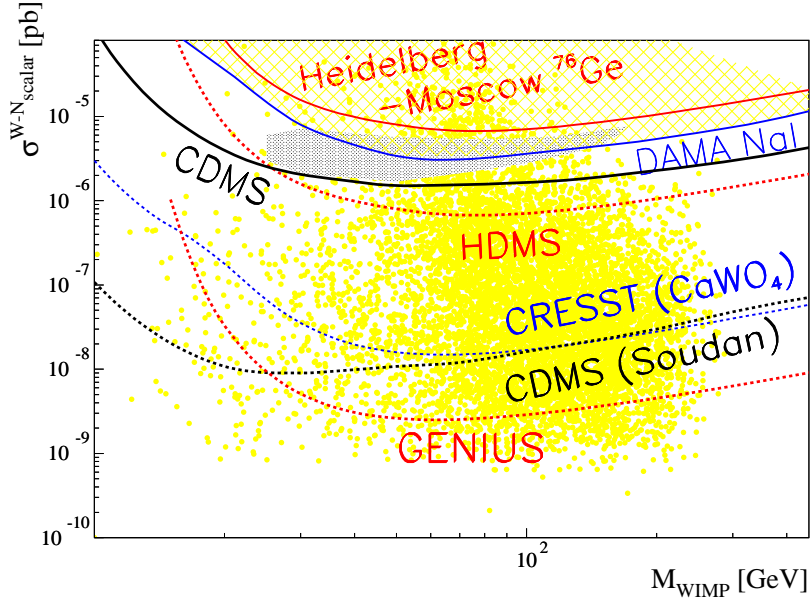}
\caption{\label{limits}
WIMP-nucleon cross section limits as a function of the WIMP
  mass for spin-independent interactions. 
The hatched region is excluded by the Heidelberg-Moscow
 \cite{prd} and the DAMA experiment \cite{dama}, the plain black
 curve is the new limit of the CDMS experiment \cite{rick2000}.
 The dashed lines are 
  expectations for recently started or future experiments, like HDMS
  \cite{hdms}, CRESST \cite{cresst}, CDMS (Soudan) \cite{cdms}
  and GENIUS \cite{prop}. The
  filled contour represents  the 2$\sigma$ evidence region of the DAMA
  experiment \cite{dama3}.    
The experimental limits are compared to
expectations (scatter plot) for WIMP-neutralinos calculated in the
MSSM parameter space at the weak scale (without any GUT constraints)
under the assumption that all superpartner masses are lower than
300 GeV - 400 GeV \cite{vadim99}. }
\end{figure}

The lowest background obtained from raw data up to now comes from 
the Heidelberg-Moscow experiment, which uses an enriched $^{76}$Ge 
detector of 2.758 kg active mass in an extreme low level environment  
in the Gran Sasso Underground Laboratory (LNGS).
After a total measuring period of more than 550 d, the background in 
the energy region 9-40 keV is about 0.05 events/kg d keV \cite{diss,prd}.
The obtained limits for the spin-independent WIMP-nucleon cross section
are very close to the present best ones (from DAMA, see below), 
and are the most stringent ones (for m$_W \geq $13 GeV) for
using only raw data (see Figure \ref{limits}).

\begin{figure}  
\centering 
\leavevmode\epsfxsize=200pt  
\epsfbox{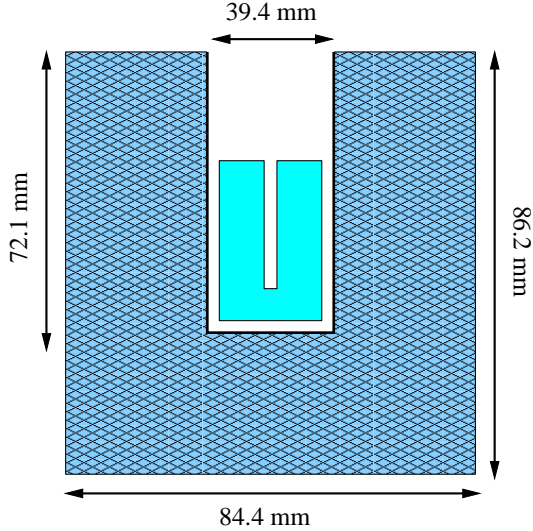}
\caption{\label{detindet} Schematic view of the HDMS experiment. A
  small Ge crystal is surrounded by a well type Ge-crystal,
  the anticoincidence between them is used to suppress background
  created by external photons.}
\end{figure}

HDMS \cite{hdms} is conceived to further reduce
the already very low background  of  the 
Heidelberg-Moscow experiment. 
 A small (200 g) Ge-crystal (natural Ge for the prototype, isotopically
enriched $^{73}$Ge for the second stage) is surrounded by a 
well-type Ge-crystal, the anticoincidence between them acting as an 
effective suppression of multiple scattered photons 
(see Figure ~\ref{detindet}). 
After a measuring period of 362.9 d, the background of the
prototype was close to the level of the Heidelberg-Moscow experiment
(the background reduction factor through anticoincidence is 4) \cite{diss}.
 After replacement of the detector holder by a low radioactive
 copper system and of the inner detector by an enriched $^{73}$Ge
crystal, the full scale experiment will start to take data in the
course of the year 2000. 

More complex Ge experiments, measuring not only the ionization, 
but also the phonon signal 
are CDMS \cite{cdms} and EDELWEISS \cite{edelweiss}. Nuclear recoils 
 deposit only about 25 \% of their energy as ionization
 (see \cite{laura-diplom} and references therein), 
while gamma - and electron interactions ionize much more 
efficiently.
Measuring  the total energy deposition provides not only the means to
achieve a low energy threshold but also a powerful background rejection
method (with exception of neutron interactions, which also induce nuclear 
recoils).
CDMS employs both thermal (the temperature change of a detector 
is measured) and athermal (the fast phonons are detected) 
phonon-mediated detectors.
Based on a 10.6 kg d exposure of thermal Ge detectors (3$\times$165~g) 
at the SUF (Stanford Underground 
Facility, 16 m.w.e.), 
new stringent limits on WIMP-nucleon scattering cross 
sections have been derived \cite{rick2000} (see Figure \ref{limits}). These exclude also 
the region allowed by the DAMA annual modulation signature 
(see below) at 84\% C.L. \cite{rick2000}.
The limitations of CDMS up to now were mainly surface events, especially electrons,
with low ionization yield (leading to confusion of these events with 
nuclear recoils). While the ionization yield of Ge detectors could 
be increased by improved charge contacts, it was realized that the fast 
phonon sensors of the Si detectors allow to 
measure the event position, and thus to identify surface events based on 
their rise-time. Recently this technique has been developed also for Ge
detectors \cite{jochen}.
CDMS will move to the Soudan mine (2070~m.w.e.), where the muon induced 
neutron background will be considerably decreased. The expected sensitivity 
at the Soudan site is shown in Figure \ref{limits}.
EDELWEISS operates two natural Ge detectors of 70~g each  in the
MODANE underground laboratory. While a test measurement with one of
the detectors resulted in a background level of 0.345~events/kg~d~keV, 
new data from the two detectors are expected soon \cite{maryvonne}.
The plans are to operate a total of 1.2~kg of Ge detectors during this year 
and, because of currently limited cryostat space, to  construct a new
cryostat system with a 100~l detector volume \cite{maryvonne}.

\begin{figure}[h!]
\centering 
\leavevmode\epsfxsize=300pt  
\epsfbox{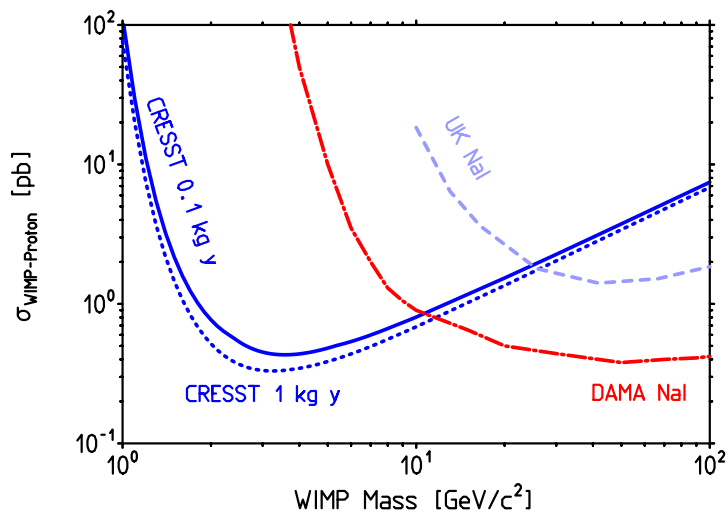}
\caption{\label{cresst_sd}
Expectation for the  WIMP-proton cross section limits
for  spin dependent interactions for the 
CRESST sapphire detectors with a total mass of 1 kg, 
an energy threshold of 0.5 keV, a background of 1\,count/(kg\,keV\,day) and
an exposure
of 0.1 and 1\,kg year. The present limits from the
DAMA
\cite{rita_sd} and UKDMC \cite{ukdm_sd} NaI experiments are also
shown (from \cite{cresst}).}
\end{figure}

Other cryogenic experiments, which however measure only the phonon signal and 
thus do not discriminate nuclear recoils are CRESST \cite{cresst} and ROSEBUD \cite{rosebud}
 (at Gran Sasso and 
Canfranc respectively,  using sapphire - Al$_2$O$_3$  - detectors),
the Milano experiment \cite{milano} (at Gran Sasso, using 
TeO$_2$ bolometers) and the Tokyo experiment \cite{tokyo} (at Nokogiryama, using LiF bolometers).
CRESST achieved the lowest energy threshold  
with 262~g sapphire detectors (500 eV) and the highest energy resolution 
(133~eV at 1.5~keV) up to now. They are thus particularly sensitive in the low 
WIMP mass region (1~GeV to 10~GeV), 
for non-coherent interactions. The expected 
sensitivity for two different exposures (0.1~kg~y and 1~kg~y) and a
 background of 1 event/kg~d~keV  is shown in Figure \ref{cresst_sd}. 
The expected sensitivity of ROSEBUD, which plans to operate 25~-~100~g of 
sapphire detectors at 2450~m.w.e. is very similar to CRESST \cite{rosebud}.

\begin{figure}[h!]  
\centering 
\leavevmode\epsfxsize=250pt  
\epsfbox{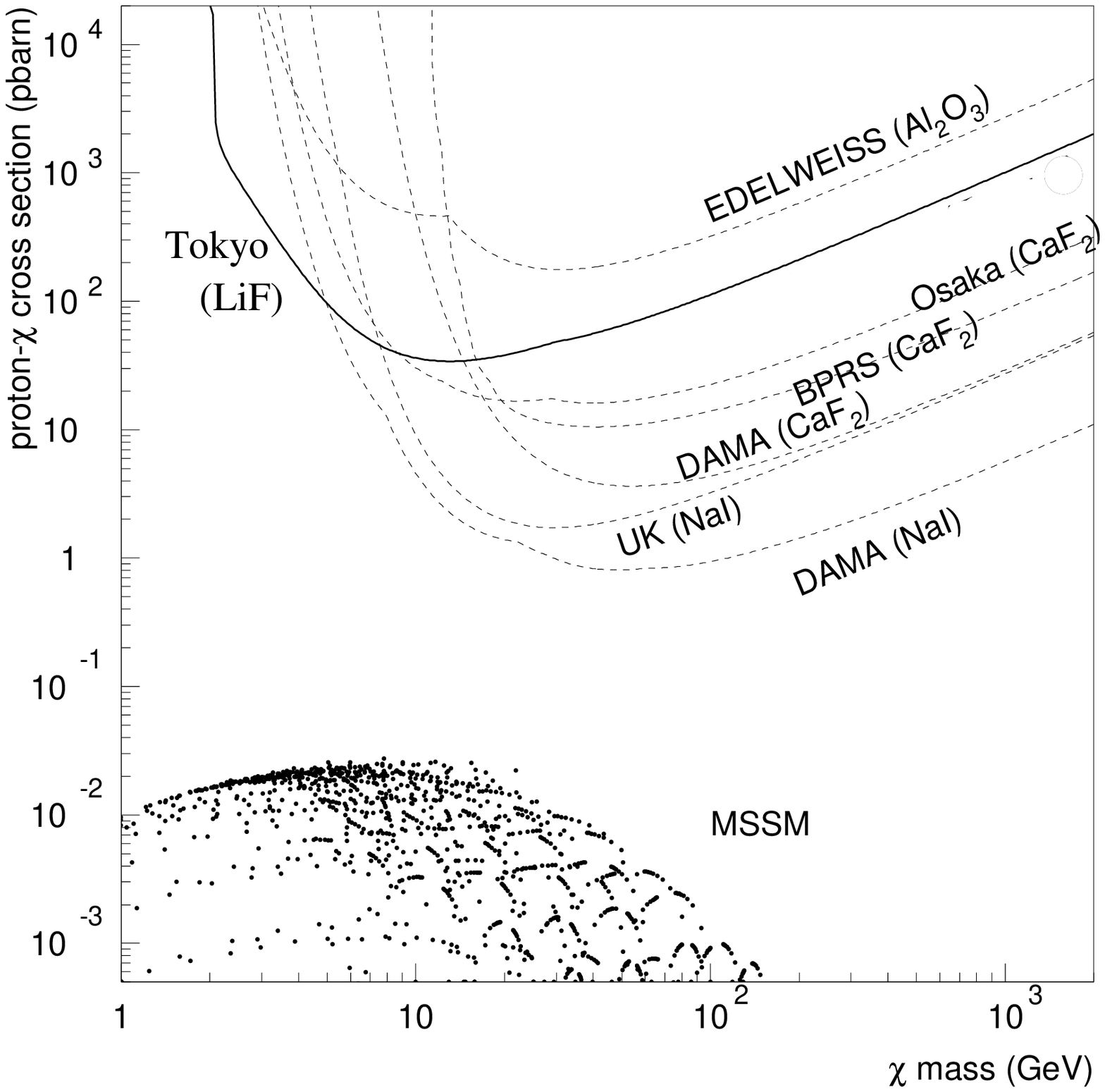}
\caption{\label{lim_sd} Summary of the  WIMP-proton cross section limits
for  spin dependent interactions, from \cite{tokyo}.}
\end{figure}

The Milano group operates an array of 20 TeO$_2$ crystals of 340 g each, 
 mainly for double beta decay searches \cite{milano}. One of the bolometers, 
with an energy threshold of 13 keV and a background level of about 
1.9 events/keV~d is used for a dark matter analysis, the limits on WIMP-proton 
cross sections however are not yet competitive to other running experiments.  
Tokyo reported results from a measurement of eight 21 g LiF bolometers 
at 15 m.w.e. for axially coupled WIMPs \cite{tokyo}. 
Although their spectra seem to be dominated 
by microphonics below 10 keV (4 detectors) and 40 keV (2 detectors), 
they improve existing limits for spin-dependent WIMP interactions for WIMP 
masses below 5 GeV (see Figure \ref{lim_sd}).

Cryogenic experiments are presently (and in the near future) limited by low 
masses. Given the complexity of the employed technology, it is
by far not trivial to operate a large amount of bolometers at mK temperatures.
The first high mass dark matter detectors were conventional scintillators 
(mostly NaI(Tl) and liquid Xenon). Their advantage is the low cost/mass-factor
(rendering masses of tens of kg feasible) and the possibility of background 
discrimination.  Although their absolute background is still at least a factor 
of ten higher than in Ge diodes, low backgrounds were achieved by severe selection 
and chemical purification of the used materials.
Background discrimination in NaI detectors relies on the shorter decay times 
of nuclear recoil than of electron induced pulses. The difference at low 
energies being small, the discrimination occurs statistically, rather than on 
an event by event basis \cite{gerbier90} (compared are the distributions of 
the decay time constants for nuclear recoils and 
for electron interactions). The pulse shape analysis requires a large 
temperature stability of an underground experiment. For example, 
in order to detect a 10\% 
component of recoils in a Compton background, the T-stability has to be better than 
1$^{\circ}$C \cite{saclay}.

Scintillator experiments which apply pulse shape analysis 
are DAMA (NaI, CaF, Xe, in Gran Sasso) 
\cite{dama},
UKDM (NaI, Xe, at the Boulby mine) \cite{ukdm} and the Saclay
experiment (NaI, in Frejus) \cite{saclay}. 

The DAMA experiment gives the at present most stringent limits on WIMP-nucleon 
cross sections (see Figures \ref{limits} and \ref{lim_sd}).  
A large target mass (about 100 kg of NaI detectors), enables DAMA to search 
for the WIMP annual modulation signature. 
In fact, an analysis of (undiscriminated) data with 
a combined exposure of 159~kg~yr is consistent with the presence of an annual 
modulation, favouring a WIMP mass of m$_W$ = (52$^{+10}_{-8}$) GeV, and a WIMP-proton 
scalar cross section of $\xi\sigma$ = (7.2$^{+0.4}_{-0.9}$)pb (where $\xi$ = $\rho_W$/0.3~GeV) 
 \cite{dama3} (see Figure 
\ref{limits} for the 2$\sigma$ confidence region).
These news have caused excitement, mainly among theorists \cite{theodama}, 
 and criticism at the same time \cite{gerbier_dama}. Being however an experimental
issue, it will most likely be solved in the near future. 

Measurements of the UKDM group with a 5 kg crystal of NaI with improved sensitivity
\cite{saclay} and of the Saclay group with a 10 kg crystal of NaI
\cite{ukdm}   revealed the existence of a small 
population of pulses with a shorter decay time even than for nuclear recoils,
with up to now unknown origin.  While the observed effect is limiting the 
pulse shape analysis method, it is currently strongly investigated.
Other limitations to pulse shape analysis in NaI detectors are 
surface interactions induced by X-rays or betas from outside the detector,
with pulse shapes very similar to nuclear recoils \cite{saclay}.

A proposal for a large liquid Xenon detector, 
with the aim to detect both scintillation and ionization signal, is the 
ZEPLIN experiment \cite{ukdm}. The 
plan is to drift the ionization electrons in a uniform electric field
and detect the current via proportional scintillation and/or induced 
electrical pulses. After test measurements with a small TPC at CERN, 
a 5 kg target mass liquid Xe detector is now under construction.
The final TPC will have a total mass of 20-30 kg.

A different type of experiments with potential to go to large detector masses 
and owning an excellent background rejection technique
 are superheated droplet detectors (PICASSO \cite{picasso}, SIMPLE \cite{simple}).
These  pure counting experiments (no energy information) use 
 small drops of a superheated liquid (freon) uniformly dispersed
in an elastic gel as detectors. While the droplets remain in a metastable,
superheated condition at ambient temperatures and pressures, an 
energy deposition by a particle with large dE/dx causes them to 
expand to bubbles after a phase transitions. The bubbles
remain fixed in the gel and can be counted as an acoustic signal by 
piezo-electric sensors.
By adjusting the operational temperature and
pressure, the droplet detectors can be made sensitive
only to nuclear recoils, the energy threshold can be varied by changing the
temperature alone.
First limits from SIMPLE (15 g active mass) and from PICASSO (5 g
active mass) for spin-dependent interaction 
are promising, although not yet competitive to other running
experiments.
The main limitations at the moment are the U/Th intrinsic background 
of the counters as well as their small active masses.

\section{Future projects}
\label{future}

The aim of future dark matter experiments lies not only in setting 
more and more stringent limits to WIMP-nucleon cross sections, 
but in eventually detecting a WIMP-specific signature, 
thus bringing an exact study of WIMP properties on a concrete ground.
The up to now proposed strategies are either to scale up existing technologies 
to large mass experiments and thus be sensitive to the predicted seasonal variation 
of the event rate or to build up a directional sensitive detector.

\begin{figure} 
\centering 
\leavevmode\epsfxsize=300pt  
\epsfbox{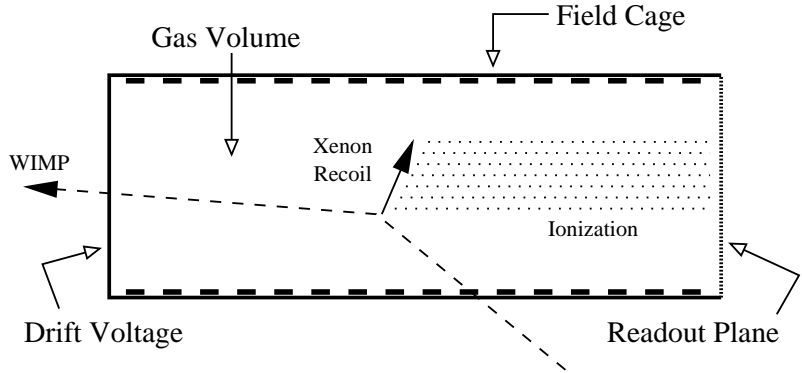}
\caption{\label{drift_tpc} Schematic view of the DRIFT TPC, from \cite{drift}.}
\end{figure}

The most sensitive directional technique proposed so far, ionization in 
a low pressure TPC, is followed by the DRIFT collaboration
\cite{drift} (see Figure \ref{drift_tpc} for a schematic view of the TPC). 
Resolving the ionization tracks in the target gas would provide not only 
a directional information but also a recoil discrimination method based on 
dE/dx and on the track length. The challenge is to minimize the electron diffusion,
in order to obtain a sufficiently high track resolution (less than 1 mm). 
While an initial small Ar test chamber (1m $\times$ 0.4 m) was placed in a 
superconducting magnet, this solution is ruled out for a ten times larger 
experiment mainly on costs reasons. A solution was found by  adding a small 
electronegative gas admixture (which would attach the ionized electrons) 
and drift the negative ions, for which both longitudinal and transversal diffusion   
 is suppressed. While small TPC prototypes are currently under 
test, the goal of the DRIFT project is to operate a 20 m$^3$ detector 
by the year 2001 in the Boulby mine \cite{drift}.

The CUORE project to be built up at Gran Sasso plans to install 1000 TeO$_2$ 
crystals (about 750 g each) in one dilution refrigerator operating at a 
temperature below 10 mK. They plan to improve their current background
level  (which is, with 1.9 events/keV d for a 340 g crystal
\cite{milano}, by a factor of about 100 higher than the one of the
Heidelberg-Moscow experiment)
by material selection and anticoincidence between the bolometers. 
It still has to be demonstrated however, that a stable low-energy threshold  
with the large TeO$_2$ crystals is feasible.

\begin{figure}[t]  
\centering 
\leavevmode\epsfxsize=250pt  
\epsfbox{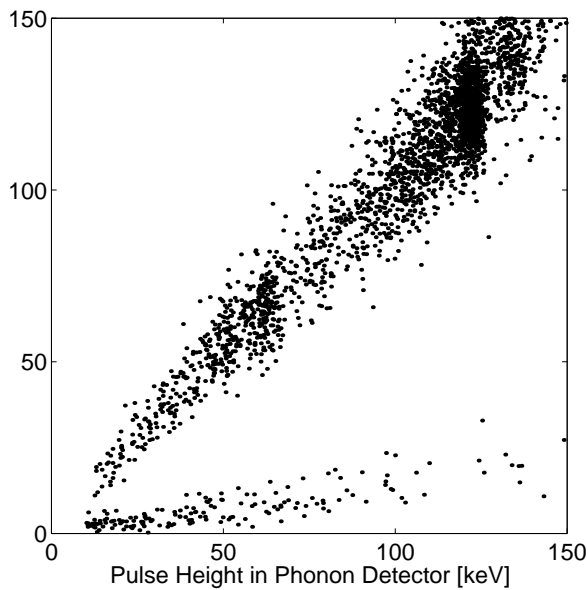}
\caption{\label{cresst_corr} Scatter plot of the pulse 
heights observed in a CRESST light detector versus
the pulse height in the phonon detector, measured with a photon 
and a neutron source. The lower, neutron induced band (nuclear
recoils) can be clearly distinguished  from the electron induced upper band.}
\end{figure}

The CRESST collaboration recently developed a nuclear recoil discrimination
method, by measuring both scintillation light and phonon signal in a CaWO$_4$ 
crystal \cite{cresst}. A small light detector is placed near the scintillating 
absorber, which has a tungsten superconducting phase transition thermometer 
on it, both being operated at 12~mK. Figure \ref{cresst_corr} shows 
the  discrimination between electron and neutron recoils 
down to an energy of 10~keV. 

While a first CaWO$_4$ crystal will be installed in Gran Sasso in the course 
of the year 2000, 
the long-term plans are to develop and install up to 100~kg of 
diverse detector materials (CaWO$_4$, PbWO$_4$, BaF, BGO) in the cold box, 
 thus to be sensitive to the annual modulation signature 
and, in case of a positive signal, to extract the WIMP properties 
through the dependence of the recoil spectrum on the target nucleus mass.
The expected sensitivity for an exposure of 100~kg~y, a background suppression 
of 99.9\% above 15 keV and a background level of 1 event/kg~d~keV is 
shown in Figure \ref{limits}.

\begin{figure}[t]  
\centering 
\leavevmode\epsfxsize=280pt  
\epsfbox{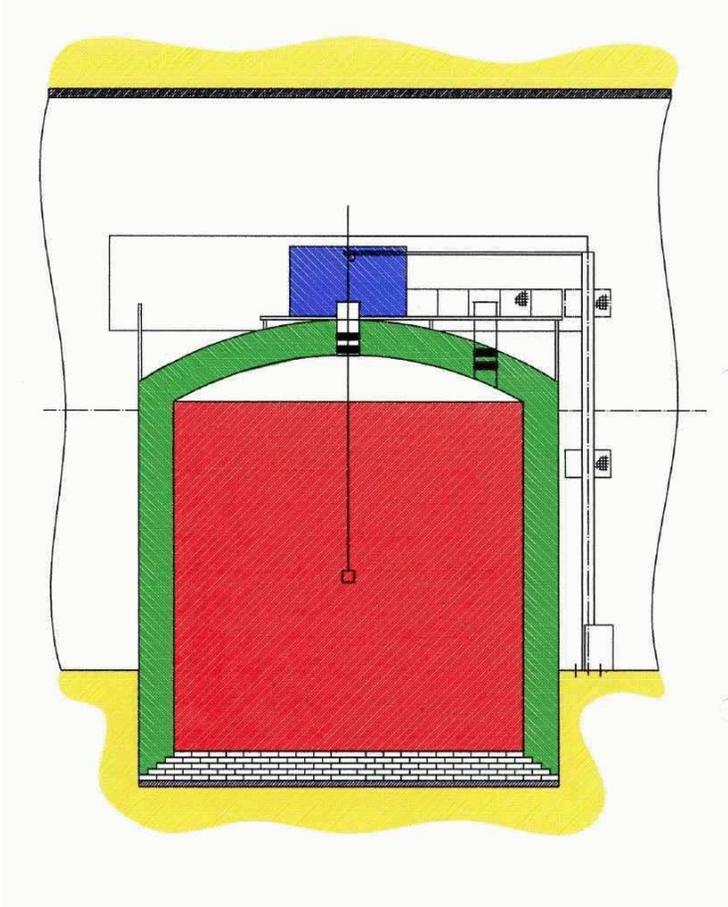}
\caption{\label{genius}Schematic view of the GENIUS tank.}
\end{figure}

For an almost complete covering of the relevant MSSM parameter space for neutralinos 
as dark matter candidates, an
increase in sensitivity by more than three orders of magnitude relative
to running experiments is required.
This is precisely the aim of the GENIUS project \cite{kla,prop}, 
which will operate about 40 
HPGe-detectors (100~kg) immersed directly in liquid nitrogen (see
Figure \ref{genius} for a schematical view).
GENIUS will use conventional ionization in a Ge crystal as detection 
technique, without discrimination of nuclear recoils.
The large step in background reduction would be achieved by removing 
(almost) all materials from the immediate vicinity of the detectors 
(crystal holder, cryostat system, which were the main background sources so 
far) and operate the crystals  directly in a cold liquid of extreme 
purity. 

GENIUS could be realized favourably in the Gran Sasso or WIPP
underground laboratories.

In three consecutive technical studies it was demonstrated that 
HPGe detectors work reliably under such conditions; low-energy thresholds 
(2.5~keV) and good energy resolutions (1~keV at 300~keV) were achieved 
with 300~g - 400~g crystals \cite{nim_genius,diss,prop}.
Monte Carlo simulations and background estimations  
showed, that achieving such a low absolute background level is feasible 
if the liquid shielding amounts to 12~m in diameter  and if low 
activation times of the crystals at sea level ($<$ 10~d) are ensured
\cite{nim_genius,diss}.
Figure \ref{limits} shows the expected sensitivity for an exposure 
of 100~kg~y and a background level of 10$^{-2}$ events/kg~y~keV.
The clear advantages of GENIUS would be the well studied detection 
technique and the possibility to increase the amount of target material 
(detectors made of enriched $^{73}$Ge material could be used to study 
the spin-dependent interaction).
Already the first step with 100~kg of natural Ge would allow not only 
to set stringent limits on WIMP-nucleon cross sections, but to 
be highly sensitive for the annual modulation signature (see also \cite{yorck_kla}).
Currently a smaller test phase, GENINO, is under study  \cite{genino}.  
GENINO would already operate 100~kg of Ge crystals, but in a smaller 
liquid nitrogen tank of 5~m diameter and 5~m height. Due to the increased influence 
of the natural environmental radioactivity, the suppression in background 
would be only a factor 20 with respect to the currently lowest value 
of about 0.05 events/kg~d~keV \cite{prd}. However, the complete 
DAMA evidence region could be tested in the near future, not only by exclusion,
but by directly looking for the modulation signature.

\section{Conclusions}

Doubtless the present situation in the field of direct dark matter 
detection is exciting.

A plenitude of experiments using very different detection techniques 
have reached sensitivities which for the first time start to probe the supersymmetric 
parameter space. A well motivated WIMP candidate 
from supersymmetry, the neutralino, might thus reveal its presence 
in the near future.
While the report of evidence for a WIMP signal by one experiment 
has raised an equal amount of interest and criticism, 
confirmation or exclusion  by other experiments is expected soon.

The goal of future projects has shifted from merely setting 
limits  on the WIMP interaction strength to the ability of detecting 
a distinctive WIMP signature. The experimentally explored WIMP signatures 
up to now, leading to unambiguous WIMP identification, are caused 
by the Earth`s motion with respect to the galactic rest frame.

The hope is that future dark matter experiments will not 
only unveil the major composition of matter in the Universe, 
but will possess the aptitude to study its inherent features, 
thus opening the avenue towards discerning between different 
galactic halo models.
Needless to say, direct detection of dark matter would have striking 
implications for particle physics and cosmology.

\section*{Acknowledgments}
We thank Y. Ramachers for a critical reading of the 
manuscript and F. Proebst and R. Golwala for 
providing informations about their experiments.


\end{document}